\documentclass{spie} 

\usepackage{graphicx}
\usepackage{lachaume}
\usepackage{subfigure}
\usepackage{notations}

\title{Bandwidth smearing in optical interferometry:\\
Analytic model of the transition to the double fringe packet.}

\author{R.~Lachaume\supit{a,b} and J.-P.~Berger\supit{c}
\skiplinehalf\small
\supit{a}~Centro de Astroingenier\'ia, Departamento de Astronom\'ia y Astrof\'isica, Pontificia Universidad Cat\'olica de Chile\\
\supit{b}~Max-Planck-Institut f\"ur Radioastronomie\\
\supit{c}~European Southern Observatory
\skiplinehalf\small
To appear in the proceedings of the 2012 SPIE Conference ``Astronomical Telescopes and Instrumentation''
}

\authorinfo{Send correspondence to R.~Lachaume, lachaume@astro.puc.cl}

\begin{document} 
\maketitle 

\begin{abstract}
Bandwidth smearing is a chromatic aberration due to the finite frequency 
bandwidth. In
long-baseline optical interferometry terms, it is when the angular extension of the source is greater than the coherence length of the interferogram.
As a consequence, separated parts of the source will contribute to fringe 
packets that are not fully overlapping; it 
is a transition from the classical interferometric regime to a double or 
multiple fringe packet. While studied in radio interferometry, there has been 
little work on the matter in the optical, where observables are measured and 
derived in a different manner, and are more strongly impacted by the 
turbulent atmosphere. We provide here the formalism and a set of usable 
equations to model and correct for the impact of smearing on the fringe 
contrast and phase, with the case of multiple stellar systems in mind.  The 
atmosphere is briefly modeled and discussed.
\end{abstract}



\section{INTRODUCTION}
\label{sec:intro}  

Long-baseline interferometry is based on the link between the complex visibility~---~the contrast and shift of the fringes~--- to the object's image, a Fourier transform\cite{VCI34,ZER38}.  Despite this straightforward relationship, the production of an image from optical interferometric data is difficult, even if a large number of measurements are recorded\cite{MON12}. One of the culprits is the turbulent atmosphere: It introduces a random variable optical delay, called piston, at each aperture, which
in turn modify the phase of the complex visibility. At radio wavelengths, the problem has been overcome either by phase referencing on a close-by point-like source\cite{DAV78,LES88}. Also self-calibration\cite{SCH80,COR81} can be used to retrieve the phases with the constraint they they cancel in a sum over telescope triplets\cite{JEN51}~---~The sum of the phases over the triplet is called the closure phase. In the optical and the IR, the anisoplanatic patch in which the atmospheric phase can be
considered uniform is too small to perform phase-referencing for most targets with the current sensitivity\cite{ESP00} and the relatively small number of telescopes (typically 3--4) make the self-calibration difficult. For this reason, optical interferometrists still predominantly fit the visibility amplitude and closure phases with parametric models\cite{MON12} with imaging being relatively recent and based on blind reconstruction techniques\cite{MAL10,BER12}.

A problem arises when interferometers are used to observe wide fields, i.e. fields that are significantly larger than their point-spread function. Because of the finite bandwidth, the van~Cittert-Zernicke theorem that links the visibility to the image is no longer exactly true and image reconstructions based on the fringes are smeared in a way that off-axis sources are elongated and less contrasted\cite{BRI89}. As seen from an optical interferometrist's point of view, this \emph{bandwidth
smearing} occurs because (i) the finite spectral resolution $R$ creates a fringe packet of finite width $R\lambda$; and (ii) the different parts of the source are separated enough, by angular distance $\theta$, for their respective fringe packets to overlap imperfectly, or even create a double fringe packet~---~the separation is $B\theta$ where $B$ is the projected baseline.  This effect becomes significant for $B\theta \sim R\lambda / 5$ \cite{ZHA07}. For a typical broad band near-IR
filter at hectometric baselines, this corresponds to 5--10 milliarcseconds of separation when a wide band filter is used without spectral resolution. The use of shorter baselines can overcome the effect without a significant precision loss on the separated parts of the source, yet this is not easily achievable nor desirable in practice, e.g. for hierarchical systems or in surveys for companions.

While this bandwidth smearing has been well modeled in radio interferometry it is still relatively difficult to overcome a posteriori in image analysis\cite{BRI89}. It is generally avoided using a higher spectral resolution. Alternatively, a posteriori software recombination with different phase centres using the amplitude and phase of the signal recorded at each antenna can solve the problem. At optical and IR wavelengths, higher resolution is generally barred by sensitivity
issues and the nature of detection prevents to record the phase of the signal for later recombination,
so the effect must be tackled. Despite the detailed description and modeling in the radio, the manner the visibility and related quantities are measured in optical and IR interferometry prevents us to use Bridle \& Schwab's formalism\cite{BRI89} as is. Since complex extended sources haven't been observed so far because of optical interferometers' limitations, we are concerned by the ``simpler'' multiple systems of near-point sources, that can be modeled using the visibility and closure phases. First steps
were taken by Zhao \textit{et al.} \cite{ZHA07} who give an approximate formula for the smearing of the visibility amplitude of a binary and numerically model that on the closure phase. However, the derivation of generic formulas and a detailed analysis of the effect have not been performed yet. 

The purpose of this paper is to provide usable formulate for the smeared observables, square visibility amplitude and closure phase, and assess the amount of bias that arises from this smearing in a practical case. We choose PIONIER\cite{PIONIER} at the Very Large Telescope Interferometer. The instrument performs pairwise coaxial recombination and delivers the interference pattern coded in temporal scans.  We finally quickly assess the influence of the atmosphere using an analytic approximation of the turbulence.

\section{Analytic description of the fringe packet}
A point-like source $o$ with flux density $\sfluxstar{o}(\xi)$ at wave number $\wavenum = \wavenumzero{0} + \xi$ is observed by an interferometer in a spectral channel centred on $\wavenumzero{0}$. We consider the injection and coupling-corrected coherent flux, i.e. an interferogram whose continuum has been removed, the instrumental contrast of the fringes has been corrected for, and the imbalance of the flux coming from both arms has been removed. For coaxial
recombination with temporal scans, it can be written as the sum over wavelengths and sources of point-source, monochromatic interferograms:
\begin{equation}
  \phasor{}{ab}(\opd{ab}(t)) 
    = \sum_o \intinf  
            \profile{}{a}(\xi) \conj{\profile{}{b}}(\xi)
            \sfluxstar{o}(\xi)
            \exp{2\mathrm{i}\pi(\wavenumzero{0} + \xi)(\xobj{o}{ab} + \opd{ab}(t) + \pist{}{ab}(t))} 
      \idiff\sigma,
\end{equation}
where 
\begin{itemize}
\item $\profile{}{a}(\xi)$ is the complex spectral transmission of the
  electric field vector of incoming radiation in the interferometer's arm, including terms that can be considered constant during a scan ($\le 1$\,s): quantum efficiency of the detector, absorption by the atmosphere, absorption along the arm, and instrumental optical path difference (OPD), i.e. instrumental phase. 
\item $\pist{}{ab}(t)$ is the atmospheric differential piston between
arms $a$ and $b$, a term that varies during a scan (at the scale of ms).
\item $\opd{ab}(t)$ is the OPD modulation on baseline $ab$.
\item $\sfluxstar{o}(\xi)$ is the flux density of source $o$.
\item $\xobj{o}{ab}$ is OPD position of the centre of the fringe packets of object $o$.  It is determined with the projected baseline $\base{ab}$ and object relative position in the sky using $\xobj{o}{ab} = \base{ab} \cdot \pos{o}$. 
\end{itemize} 
It can be rewritten as
\begin{align}
  \phasor{}{ab}(\opd{ab}(t)) 
    &= \sum_o 
            \underbrace{\strut\IFT{
                \profile{}{a} \conj{\profile{}{b}}
                \sfluxstar{o}
            }(\xobj{o}{ab} + \opd{ab} + \pist{}{ab}(t))}_{\text{Envelope $+$ instrumental phase}}
            \underbrace{\exp{\mathrm{i}\phiobj{o}{ab} 
            + 2i\pi\wavenumzero{0}(\opd{ab} + \pist{}{ab}(t))}\strut}_{\text{Fringes}}, 
\end{align}
where 
\begin{itemize}
\item $\phiobj{o}{ab} = 2\pi\wavenumzero{0}\xobj{o}{ab}$ is the fringe packet shift in terms of phase.
\item $\IFT{f}$ is the inverse Fourier transform of function $f$.
\end{itemize}

\section{Piston-free observables}
In that case the coherent flux and its Fourier transform, the spectral density, read: 
\begin{align}
  \phasor{}{ab}(\opd{ab}) 
    &= \sum_o 
            \IFT{
                \profile{}{a} \conj{\profile{}{b}}
                \sfluxstar{o}
            }
            (\xobj{o}{ab} + \opd{ab})
        \,
        \exp{   \mathrm{i}\phiobj{o}{ab} 
        + 2\mathrm{i}\pi\wavenumzero{0}\opd{ab}},
  \label{eq:def:phasor}
\\
  \specdens{}{ab} (\wavenum) &= 
         \sum_o  
           \profile{}{a}\conj{\profile{}{b}}\sfluxstar{o}
           (\wavenum - \wavenumzero{0}) 
           \,\exp{2i\pi\wavenum\xobj{o}{ab}},
  \label{eq:def:specdens}
\end{align}
\begin{figure}
\centering
\includegraphics[width=0.85\textwidth]{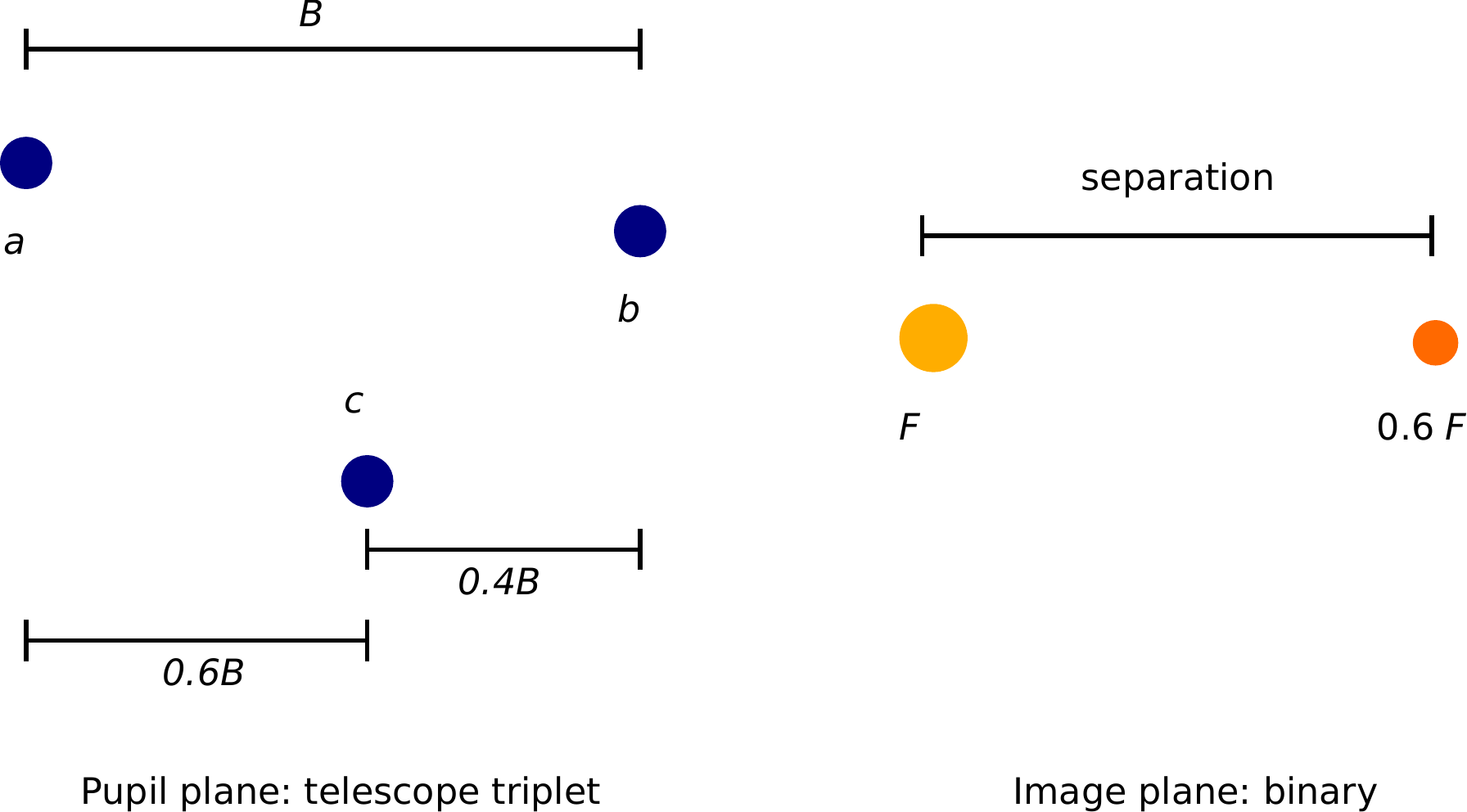}
\caption{The telescope triplet and binary given in our test case.  The telescope triplet $a$, $b$, $c$ features baseline lengths $B$, $0.6B$, and $0.4B$ along the orientation of the binary, of flux ratio 0.6.}
\label{fig:bin}
\end{figure}

\begin{figure}[p]
\centering
\includegraphics[width=0.85\textwidth]{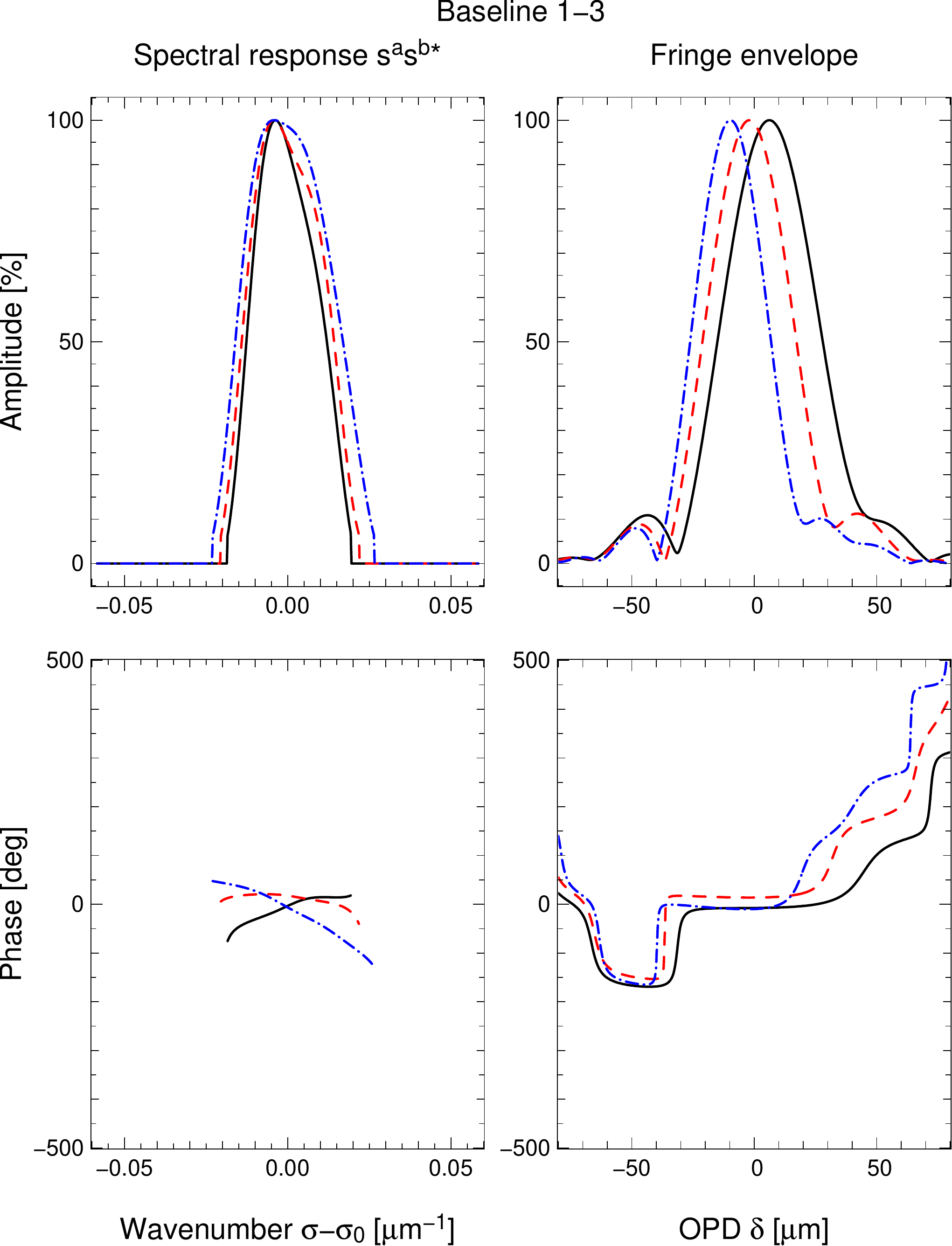}
\caption{Example of transmission for PIONIER. The internal lamp was observed on {2011-08-09} in the H band with the 3-channel spectral resolution, a sample baseline is reported here. The left columns display the spectral transmission and phase ($\profile{}{a}\conj{\profile{}{b}}$) for the three channels (from larger to lower wavelength: black \& solid, red \& dashed, and blue \& dashed-dotted lines).  The right columns shows the envelope and phase of the fringe packet (the Fourier
transform of the latter). The phase slope of the spectral response (bottom left) translates into a shift of fringe packet centering, known as group delay (top right).}
\label{fig:PT}
\end{figure}

\begin{figure}
\subfigure[Interferogram at maximal binary separation (25~$\text{mas}\cdot\text{hm}\cdot\mu\text{m}^{-1}$)]
{\includegraphics[width=\textwidth]{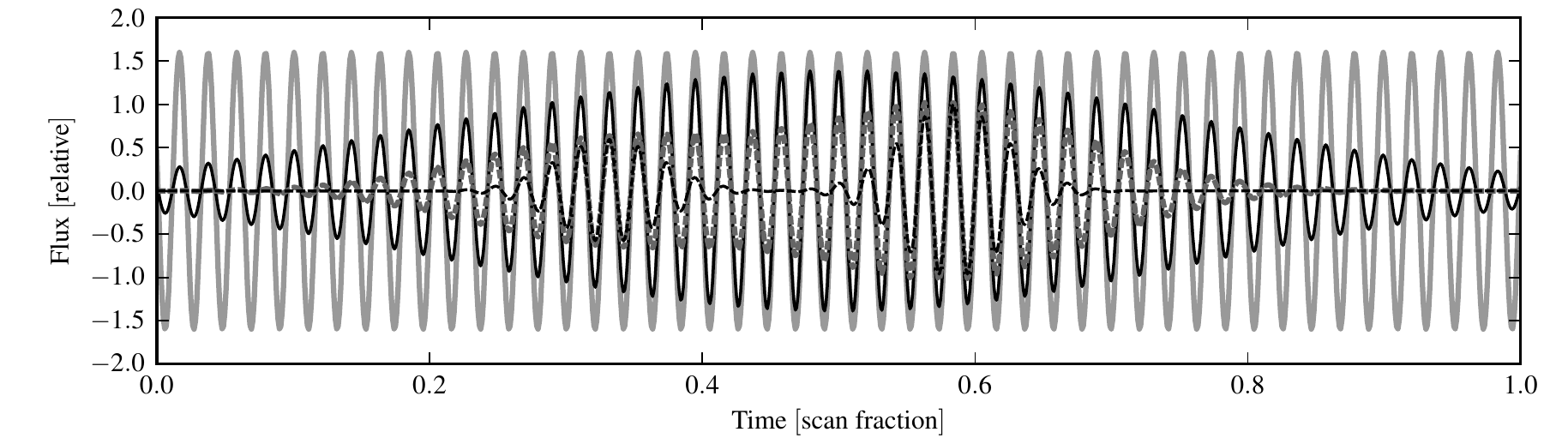}
\label{subfig:I}
}\\
\subfigure[Visibility amplitude as a function of binary separation]
{\includegraphics[width=\textwidth]{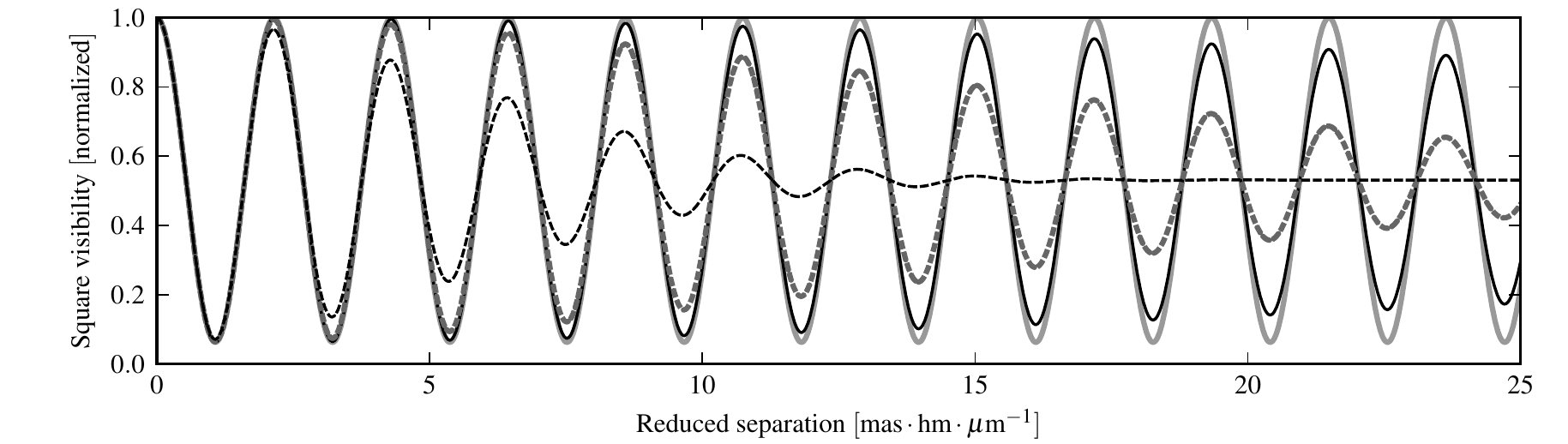}
\label{subfig:V}
}\\
\subfigure[Closure phase as a function of binary separation]
{\includegraphics[width=\textwidth]{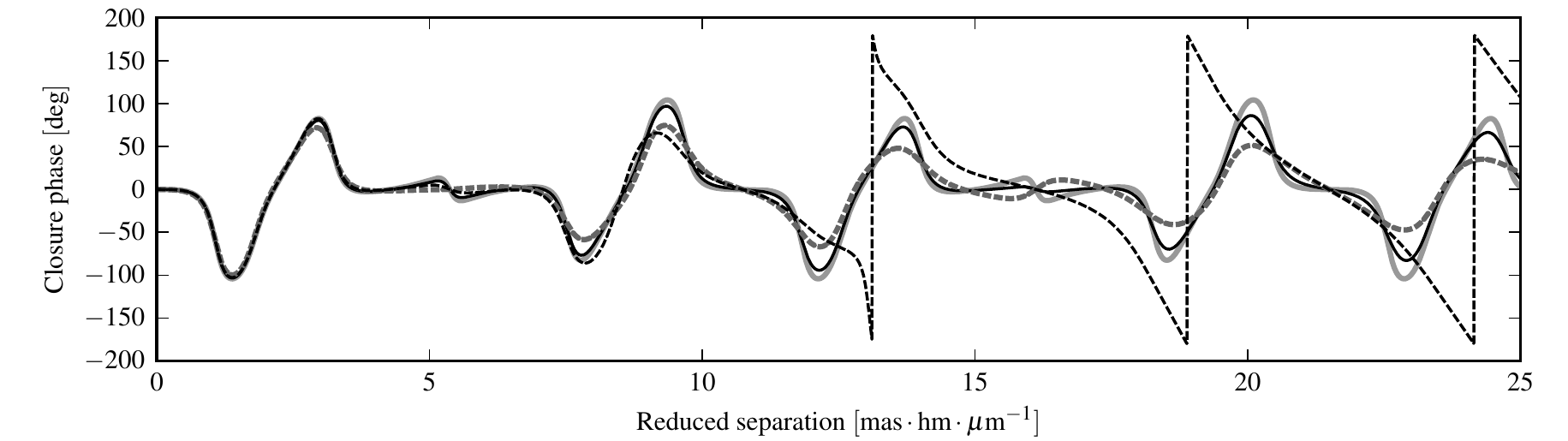}
\label{subfig:phi}
}
\caption{Example of a binary observed with an ideal interferometer of infinite  spectral resolution (gray solid line) and with the three main spectral          resolutions of PIONIER in the H band (7 channels, dark solid line; 3, gray      dashed line; and whole band, dark dark line). 
The lines are easier to distinguish by level of smearing, the further from the ideal solid curve, the smaller the resolution.  For the visibility and phase, the reduced binary separation is expressed in milliarcseconds-hectometres per micron for the sake of generality, i.e. in milliarcseconds for a hectometric projected baseline at a wavelength of one micron.  
For the closure phase the triplet of telescopes is taken with coordinates (0, $B$, $0.4B$) along the binary separation axis and the largest baseline $B$ is used in the horizontal axis.  In the whole band setting, the group delay is zero; in other settings we assumed a spectral channel with (-10, 0, 10) $\mu$m group delay.  It corresponds to our PIONIER test case, shown in Fig.~\ref{fig:PT}.}
\label{fig:pionier}
\end{figure}

It is possible to measure $\profile{}{a}\conj{\profile{}{b}}$ on fringes
obtained on a calibration lamp, provided that its spectrum is known, 
because piston is almost zero.  We did it in the case of PIONIER, an example is shown in Fig.~\ref{fig:PT}.  Over the primary lobe of the fringe packet envelope the main trends seen on Fig.~\ref{fig:PT} are:
\begin{itemize}
\item A constant instrumental phase over the primary lobe of the fringe packet (compare the flat parts of the curves in bottom right panel to the fringe envelope in the upper right one).
\item A clear group delay trend, with fringe packets centered at different positions in the different channels (top right panel). It is produced by the chromaticity of the instrumental differential phase (bottom left panel). 
\end{itemize} 

In the following, we will approximate this by using a Gaussian profile of the spectral transmission, with a group-delay $\gd{ab}$ and constant instrumental phase $\insphi{}{ab}$:
\begin{align}
  \profile{}{a}\conj{\profile{}{b}}(\xi) &= 
    \wideexp{\left[
       -\frac{2 \log 2 \, \xi^2}{\dwavenum{}^2}
       +\mathrm{i}\left( 2\pi\gd{ab}\xi + \insphi{}{ab}\right)
    \right]},
    \label{eq:sasb:approx}\\
  \IFT{\profile{}{a}\conj{\profile{}{b}}}(\delta) &= 
    \wideexp{
        \left[
           -\frac{\pi^2\dwavenum{}^2(\delta-\gd{ab})^2}{2\log 2}
           + \mathrm{i}\insphi{}{ab}
        \right]
      }.
    \label{eq:tfsasb:approx}
\end{align}

Within this approximation, we take the example of a binary with flux ratio 0.6, and show its interferogram in Fig.~\ref{subfig:I} for the three spectral resolutions of PIONIER in the H band (1, 3, 7 channels, or $\resol{} \approx 5, 15, 35$) as well as the ideal interferometer are given in Fig.~\ref{subfig:I}.  See Fig.~\ref{fig:bin} for a sketch of the interferometer's configuration.

\subsection{Visibility}
The visibility amplitude can be determined using
\begin{equation}
  |V|^2 = \frac1{\normtot{}} \intinf \phasor{}{ab}(\opdvar)
                          \conj{\phasor{}{ab}}(\opdvar) 
                          \idiff\opdvar
        = \frac1{\normtot{}} \intinf \specdens{}{ab}(\sigma) 
                          \conj{\specdens{}{ab}}(\sigma)
                          \idiff\sigma 
\end{equation}
where $\normtot{}$ is a convenient normalisation factor.  From Eq.~(\ref{eq:def:specdens}),
\begin{equation}
  |V|^2 = \frac1{\normtot{}} \sum_{o, p} 
                 \IFT{\sfluxtel{o}{a}\sfluxtel{p}{b}}
                    (\xobj{o}{ab} - \xobj{p}{ab})
                 \,
                 \exp{\mathrm{i}(\phiobj{o}{ab} - \phiobj{p}{ab})}.
\end{equation}

We now consider the case of the Gaussian transmission, assuming sources with a flat spectrum $\sfluxstar{o}$. The group-delay term of Eq.~(\ref{eq:sasb:approx}) cancels out in the visibility amplitude formula so that:
\begin{equation}
   |V|^2 =
        \sum_o \frac{\flux{o}^2}{\fluxtot{}^2} 
        + \sum_{o < p}
        \frac{2\flux{o}\flux{p}}{\fluxtot{}^2}
        \cos \left(\phiobj{o}{ab}-\phiobj{p}{ab}\right) 
        \underbrace{
          \wideexp{
            -\frac{(\phiobj{o}{ab}-\phiobj{p}{ab})^2}{16\resol{}^2\log 2}
          }
        }_{\text{smearing}}
        , 
\label{eq:sameprofile:vsqPS}
\end{equation}
where $\resol{} = \wavenumzero{0} / \dwavenum{}$ is the spectral resolution.

Within this approximation, the visibility of a binary with flux ratio 0.6 (see Fig.~\ref{fig:bin}, baseline $ab$) for the three spectral resolutions of PIONIER in the H band (1, 3, 7 channels, or $\resol{} \approx 5, 15, 35$) as well as the ideal visibility are given in Fig.~\ref{subfig:V}.  

\subsection{Closure phase}
The closure phase is the argument of the bispectrum, determined in direct space using
\begin{equation}
  B^{abc} = \intinf 
                    \phasor{o}{ab}(\opd{ab}(t)) 
                    \phasor{p}{bc}(\opd{bc}(t)) 
                    \phasor{q}{ca}(\opd{ca}(t))
                    |\dopd{ab}(t)\dopd{bc}(t)\dopd{ca}(t)|
                    \idiff t,
  \label{eq:def:bispDS}
\end{equation}
In the case of linear OPD variations we write $\opd{ab}(t) = \dopd{ab} t$ conveniently choosing the origin of time.  It follows from Eqs.~(\ref{eq:def:phasor}~\& \ref{eq:def:bispDS}) and closure relation $\dopd{ab} + \dopd{bc} + \dopd{ca} =   0$ that
\def\DIRsflux#1#2#3#4{\profile{}{#3}
                      \conj{\profile{}{#4}}
                      \sfluxstar{#1}}
\def\IFTsflux#1#2#3#4{\IFT{\DIRsflux#1#2#3#4}}
\begin{equation}
  B^{abc} =
    \sum_{o, p, q} 
    \exp{\mathrm{i}(\phiobj{o}{ab} + \phiobj{p}{bc} + \phiobj{q}{ca})}
     \intinf 
             \IFTsflux olab(\xobj{o}{ab} + \dopd{ab} t)
             \IFTsflux plbc(\xobj{p}{bc} + \dopd{bc} t)
             \IFTsflux qlca(\xobj{q}{ca} + \dopd{ca} t)
             \idiff t.
\label{eq:int:bispDS}
\end{equation}
It can be identified as a triple correlation and equated to the two-dimensional Fourier transform of a product:
\begin{equation}
  B^{abc} \propto\sum_{o, p, q} 
    \IFT{\striple{opq}{ab}}
      \left(
        \xobj{o}{ab} - \frac{\dopd{ab}}{\dopd{ca}} \xobj{q}{ca}, 
        \xobj{p}{bc} - \frac{\dopd{bc}}{\dopd{ca}} \xobj{q}{ca}
      \right) \,
      \exp{\mathrm{i}\left( \phiobj{o}{ab} + \phiobj{p}{bc} + \phiobj{q}{ca}   \right)}.
\label{eq:gen:bispDS}
\end{equation}
with the triple spectral response
\begin{equation}
  \striple{opq}{ab}(\xi_1, \xi_2) = 
     (\DIRsflux olab)(\xi_1) \, 
     (\DIRsflux plbc)(\xi_2) \,
     (\DIRsflux qlca) 
     \left(
        - \frac{\dopd{ab} \xi_1 + \dopd{bc} \xi_2}{\dopd{ac}}
     \right).
  \label{eq:def:striple}
\end{equation}

In the case of a Gaussian throughput with linear group-delay (Eq.~\ref{eq:sasb:approx}) and a flat source spectrum:
\def\phiobjgd#1#2{\phiobj{#1}{#2\prime}}
\begin{equation}
  B^{abc} \propto 
  \exp{\mathrm{i}\insphi{}{abc}}
      \sum_{o,p,q} 
    \flux{o}\flux{p}\flux{q}
    \exp{\mathrm{i}\left( \phiobj{o}{ab} + \phiobj{p}{bc} + \phiobj{q}{ca} \right)}
    \underbrace{\exp{
      -\frac{
          \left(\dopd{bc}\phiobjgd{o}{ab} - \dopd{ab}\phiobjgd{p}{bc}\right)^2
        + \left(\dopd{ca}\phiobjgd{p}{bc} - \dopd{bc}\phiobjgd{q}{ca}\right)^2
        + \left(\dopd{ab}\phiobjgd{q}{ca} - \dopd{ca}\phiobjgd{o}{ab}\right)^2
       }{8\resol{}^2\log2((\dopd{ab})^2 + (\dopd{bc})^2 + (\dopd{ca})^2)}}
    }_{\text{Smearing}},
    \label{eq:gauss:bispDS}
\end{equation}
where $\insphi{}{abc} = \insphi{}{ab} + \insphi{}{bc} + \insphi{}{ca}$ is the instrumental closure phase and $\phiobjgd{o}{ab} = \phiobj{o}{ab} -  2\pi\wavenumzero{0}\gd{ab}$ is the group-delay-corrected position of the fringe system for object $o$ on base $ab$.  It is worth noting that even with small separations ($\phiobj{o}{ab} \approx 0$) the smearing term is non zero because of the group delay.  Our example of a binary is display in Fig.~\ref{subfig:phi}. 

\section{Modelling the atmosphere}
The extraction of the visibility amplitude and closure phase are meant to be robust against atmospheric turbulence, that is, they should not be biased.  However, this has only been studied for compact sources where no smearing is present.  We here quickly assess the combined influence of atmosphere and smearing.

An analytic approach to the atmospheric turbulence can be taken in the
slow turbulence regime, using the assumption that scanning is fast enough 
for the piston to vary linearly during one scan, i.e.
$\pist{}{ab} = \pist{0}{ab} + \pist{1}{ab} \opd{ab}$, where $\pist{0}{ab}$
is the group-delay tracking error and $\pist{1}{ab}$ a rate of piston variation
during scan. $\pist{0}{ab}$ and $\pist{1}{ab}$ are random variables of zero 
mean.   

Using this approach,
\begin{align}
  \phasor{}{ab}(\opd{ab}) 
    &= \sum_o 
            \IFT{
                \profile{}{a} \conj{\profile{}{b}}
                {\sfluxstar{o}}^2
            }(\xobj{o}{ab} + \pist{0}{ab} + (1+\pist{1}{ab})\opd{ab})
        \,
       \exp{    
              2i\pi\wavenumzero{0}(\xobj{o}{ab} + \pist{0}{ab})
            + 2i\pi\wavenumzero{0}(1 + \pist{1}{ab})\opd{ab}},
\label{eq:pist:phasor}
\\
  \specdens{}{ab} (\wavenum) &= 
         \sum_o  
           \profile{}{a}\conj{\profile{}{b}}\sfluxstar{o}
               \left(
                 \frac{\wavenum}{1+\pist{1}{ab}} - \wavenumzero{0}
               \right)
           \,\exp{2i\pi\frac{\wavenum}{1 + \pist{1}{ab}}
                   (\xobj{o}{ab} + \pist{0}{ab})},
  \label{eq:pist:specdens}
\end{align}

\subsection{Visibility.} The determination of the visibility is straightforward.  The coherent norm
stays identical, while the power spectrum in Fourier space involves a 
contraction of factor $1 + \pist{1}{ab}$ of $\specdens{l}{ab}$. Thus
the only difference is a factor independent of separation (and smearing):
\begin{equation}
  |V^2|_{\text{jitter}} = \frac{1}{1+\pist{1}{ab}} |V^2|_{\text{ideal}} 
\end{equation}
As long as $\pist{1}{ab} \ll 1$, the linear approximation ensures that it cancels out on a large number of scans.

\subsection{Closure phase in direct space.} 
The piston terms cancel over a baseline triplet, so that the only changes 
concerns the triple correlation of the spectral response. The constant 
piston term can be seen as a fringe shift that adds 
to the original fringe shift of the object and the group delay 
\begin{equation*}
  \xobj{o}{ab} + \gd{ab} \rightarrow \xobj{o}{ab} + \gd{ab} + \pist{0}{ab}
\end{equation*}
and the linear variation of the piston can be seen as a scanning velocity 
change
\begin{equation*}
  \dopd{ab}  \rightarrow \dopd{ab} + \pist{1}{ab}
\end{equation*}  
The jittered bispectrum can thus be obtained from the piston-free formulas 
(Eqs.~\ref{eq:gen:bispDS}, \ref{eq:def:striple}~\& \ref{eq:gauss:bispDS}) by performing these substitutions. 

It can be shown that a concave smearing function---as it is the case for the exponential in our approximation Eq.~\ref{eq:gauss:bispDS}--- will tend to give a lesser value (i.e. more smearing) when averaged over zero mean random variables $\pist{i}{ab}$.  We therefore expect that the atmosphere increases the smearing effects where they are already present.  This is both confirmed in a work done by Zhao et al.\cite{ZHA07} with the Infrared Optical Telescope Array (but with a different
data reduction method, thus bispectrum formula) and by a full numerical simulation of the atmosphere and interferometer that we are developing.

\section{Conclusion}
The smearing of the visibility and the closure phase derived from temporal scans can be modeled from a spectral calibration of the interferometer.  We gave exact, generic analytic formulas and an application to a Gaussian band pass.  This can be applied to ``wide'' multiple systems.  When the components of the system are resolved analytic formulas can also be derived though they are slightly more complicated and do not bring insight into the phenomenon of
smearing itself. 

The turbulent atmosphere has an impact on the amount of smearing of closure phase, though.  While a reasonable analytic approximation can be done for each scan, there is no obvious method of taking atmosphere into account other than its average properties.  Averaging the smeared observables over scans in the presence of the atmosphere needs some numerical modeling of the random atmosphere, which we are now undertaking. 

\appendix

\section{Some questions raised during the conference}

\paragraph{Is there an simple explanation of where the smearing comes from?}  It
comes from the finite frequency band width. With infinite spectral resolution there is no smearing because fringes packets are infinitely long.  To understand the phenomenon in optical interferometric terms, one can consider the extreme case of the double fringe packet: the fringes of one component are diluted by the incoherent flux of the other component, and the other way round, which leads to a visibility loss.  

\paragraph{What differs from the radio?}  We are measuring a square visibility
amplitude instead of a complex visibility; the observables can be extracted from
the interferogram in different, non equivalent manners; the atmosphere varies
significantly during a scan.  Formalisms look very similary, but there are significant differences in the formulae.

\paragraph{Does it work for point sources only?}  It's possible to have compact
resolved sources (i.e. each source, taken separately, is unsmeared) with little additional formalism.  Also, sums can be substituted by integrals for more generality.

\paragraph{Is it model dependent?} Yes, you have to know what you source look like to model the smearing with this method.  Also, you should know the spectrum of all parts of the source to accurately determine the fringe packet envelopes with accuracy; in the near IR, assuming a flat spectrum is a good enough approximation though (at least for young stellar objects).

\paragraph{Why are there zeroes of the atmospheric bias in the smearing of the closure phase?} I couldn't figure out the reason of this finding by Zhao et al.\cite{ZHA07} in the analytic formulae.

\begin{acknowledgements}
This research has made use of NASA's Abstract Data System, the free softwares maxima, Yorick, and python. It has been supported by Comit\'e Mixto ESO-Chile and Basal-CATA (PFB-06/2007). 
\end{acknowledgements}

\bibliography{biblio}
\bibliographystyle{spiebib}   

\end{document}